\documentclass{msf}

\def\beq{\begin{equation}}
\def\eeq{\end{equation}}
\def\qvec{{\bf q}}
\def\qhat{\hat{\bf q}}

\def\etal{{\it et al.\ }}

\def\epsbol{\mbox{\boldmath$\epsilon$}}
\def\sigbol{\mbox{\boldmath$\sigma$}}

\def\mn{m_{\scriptscriptstyle N}}
\def\N{{\scriptscriptstyle N}}
\def\A{{\scriptscriptstyle A}}

\def\PRC{{\em Phys. Rev.} C}
\def\PR{{\em Phys. Rev.}}
\def\IJMPE{{\em Int.\ J. Mod.\ Phys.}\ E}
\def\CzJP{{\em Czech.\ J. Phys.}}

\begin{document}

\title{Spin polarisabilities of the nucleon at NLO in the chiral expansion}

\author {Judith A. McGovern and Michael C. Birse}
\address{Theoretical Physics Group, Department of Physics and Astronomy\\
University of Manchester, Manchester, M13 9PL, U.K.
\\ E-mail: judith.mcgovern@man.ac.uk}

\author{K. B. Vijaya Kumar}
\address{Department of Physics,
University of Mangalore, Mangalore 574 199, India}
\maketitle
\abstracts{
We present a calculation of the fourth-order (NLO) contribution to 
spin-dependent Compton scattering in heavy-baryon chiral perturbation 
theory, and we give results for the four spin polarisabilities. No low-energy
constants, except for the anomalous magnetic moments of the nucleon, enter
at this order.  The NLO contributions are as large or larger than the LO pieces,
making comparison with experimental determinations questionable.  We address
the issue of whether one-particle reducible graphs in the heavy baryon theory
contribute to the polarisabilities.}

\section{Introduction}

The usual notation for the Compton scattering amplitude in
the Breit frame is, for incoming real photons of energy $\omega$ and
momentum $\qvec$ to outgoing real photons of the same energy energy and
momentum $\qvec'$,
\begin{eqnarray}
T&=&\epsilon'^\mu\Theta_{\mu\nu} \epsilon^\nu\nonumber\\
&=&\epsbol'\cdot\epsbol\,A_1(\omega,\theta) 
+\epsbol'\cdot\qhat\,\epsbol\cdot\qhat'\,A_2(\omega,\theta) \nonumber\\
&&+i\sigbol\cdot(\epsbol'\times\epsbol)\,A_3(\omega,\theta)+
i\sigbol\cdot(\qhat'\times \qhat)\,\epsbol'\cdot\epsbol \,
             A_4(\omega,\theta)\nonumber\\
&&+\Bigl(i\sigbol\cdot(\epsbol'\times \qhat)\,\epsbol\cdot\qhat'-
i\sigbol\cdot(\epsbol\times \qhat')\,\epsbol'\cdot\qhat\Bigr)\,
             A_5(\omega,\theta)\nonumber\\
&&+\Bigl(i\sigbol\cdot(\epsbol'\times \qhat')\,\epsbol\cdot\qhat'-
i\sigbol\cdot(\epsbol\times \qhat)\,\epsbol'\cdot\qhat\Bigr)\,
             A_6(\omega,\theta),
\label{amp}
\end{eqnarray}
where hats indicate unit vectors.
By crossing symmetry the functions $A_i$ are even in $\omega$ for $i=1,2$ and
odd for $i=3-6$.  The leading 
pieces in an expansion in powers of $\omega$ are given by low-energy
theorems\cite{LGG}, and the next terms contain the electric and magnetic
polarisabilities $\alpha$ and $\beta$ and the spin polarisabilities
$\gamma_i$:
\begin{eqnarray}
A_1(\omega,\theta)\!&=&\!-{Q^2\over\mn}
%-{Q^2\over 4\mn^3}\omega^2(1-\cos\theta)
+4\pi(\alpha+\cos\theta\beta)\omega^2+{\cal O}(\omega^4)\nonumber\\
A_2(\omega,\theta)\!&=&\!-4\pi\beta\omega^2+{\cal O}(\omega^4)\nonumber\\
A_3(\omega,\theta)\!&=&\!
{e^2 \omega\over 2\mn^2}\Bigl(Q(Q+2\kappa)-(Q+\kappa)^2 \cos\theta\Bigr)
+4\pi\omega^3(\gamma_1+\gamma_5\cos\theta)
%\nonumber\\ &&-{e^2 Q(Q+2\kappa)\omega^3\over 8\mn^4}
+{\cal O}(\omega^5)\nonumber\\
A_4(\omega,\theta)\!&=&\! -{e^2\omega \over 2\mn^2 }(Q+\kappa)^2 
+4\pi\omega^3 \gamma_2 +{\cal O}(\omega^5)\nonumber\\
A_5(\omega,\theta)\!&=&\! {e^2 \omega\over 2\mn^2 }(Q+\kappa)^2 
+4\pi\omega^3\gamma_4 +{\cal O}(\omega^5)\nonumber\\
A_6(\omega,\theta)\!&=&\! -{e^2 \omega\over 2\mn^2 }Q(Q+\kappa)
+4\pi\omega^3\gamma_3 +{\cal O}(\omega^5)
\end{eqnarray}
where the charge of nucleon is $Q=(1+\tau_3)/2$ and its anomalous magnetic 
moment is $\kappa=(\kappa_s+\kappa_v\tau_3)/2$.
Only four of the spin polarisabilities are independent since three are related
by  $\gamma_5+\gamma_2+2\gamma_4=0$.  The polarisabilities are isospin 
dependent.

Compton scattering from the nucleon has recently been the subject of much
work, both experimental and theoretical. The unpolarised polarisabilities have
been well known for a number of years now, at least for the neutron, but it
is only very recently that determinations of the spin polarisabilities have been
extracted from fixed-$t$ dispersion analyses of photoproduction data.
The forward spin polarisability $\gamma_0=\gamma_1+\gamma_5$ has a longer history,
with determinations that are in the range of recent values, namely 
$-0.6$ to $-1.5\times 10^{-4}$~fm$^4$ for the proton.\cite{sand,krein2,babusci,DGPV}
Direct measurements of the polarised cross-section at MAMI have been used to
obtain a value of $-0.8\times 10^{-4}$~fm$^4$, as reported by Pedroni at this 
conference.
No direct measurements of polarised Compton scattering have yet been
attempted.  However the backwards spin polarisability 
$\gamma_\pi=\gamma_1-\gamma_5$ has recently been extracted from unpolarised  
Compton scattering from the proton.  The LEGS group\cite{tonnison} obtained 
$-27\times 10^{-4}$~fm$^4$, far from the previously accepted value of 
$-37\times 10^{-4}$~fm$^4$, which is dominated by $t$-channel pion exchange.
In contrast results presented by Wissmann at this conference give a value
extracted from TAPS data which is compatible with the old value.

\section{Polarisabilities in HBCPT}

The non-spin polarisabilities have previously been determined to NLO (fourth order)
in heavy baryon chiral perturbation theory (HBCPT).  The values are in
good agreement with experiment, with the NLO contribution (where LEC's enter)
being small compared to the LO part (which comes from pion-nucleon loops).
The spin polarisabilities have also been calculated;\cite{mei95} 
at lowest order the value
$\gamma_0=\alpha_{em}g_\A^2/(24 \pi^2 f_\pi^2 m_\pi^2)=4.51$ is obtained 
for both proton and neutron, where the entire contribution comes from  $\pi N$ 
loops.\cite{ber92} The
effect of the $\Delta$ enters in counter-terms at fifth  order in standard
HBCPT, and has been estimated to be so large as to change the sign\cite{ber92}. 
The calculation has also been done in an extension of HBCPT
with an explicit  $\Delta$ by Hemmert {\it et al.}\cite{hemm}  They find that the
principal effect is from the $\Delta$ pole, which contributes $-2.4$, with
the effect of $\pi\Delta$ loops being small, $-0.2$.  Clearly the next most
important contribution is likely to be the fourth-order $\pi N$ piece, and
this is the result which is presented  here.\cite{kumar}

\begin{figure}[t]
%\figurebox{20pc}{15pc}{} % to have a box alone
\epsfxsize=27pc % will enlarge or reduce the postscript figures based on the xsize
\epsfbox{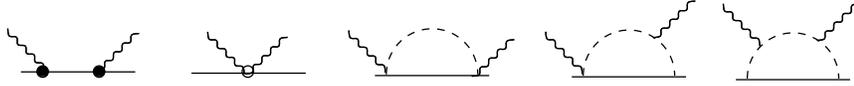} % postscript image file name
\caption{Diagrams which contribute to spin-dependent Compton scattering in the 
$\epsilon\cdot v=0$ gauge at LO. The solid dots are vertices
from ${\cal L}^{(2)}$ and the open circle is a vertex from ${\cal L}^{(3)}$.}
\end{figure}

Two other groups have also presented fourth order calculations of the 
spin polarisabilities recently; Ji \etal calculated $\gamma_0$ and obtained
an expression in complete agreement with ours.\cite{osborne} Gellas \etal have also
calculated all four polarisabilities.\cite{gellas} 
Their calculations agree with ours,
but we disagree on what constitutes the polarisabilities; we will say more
about this later.

In HBCPT the fixed terms in the amplitudes $A_3$ to $A_6$ are reproduced at
leading (third) order, by the combination of the Born terms and the seagull
diagram. The same terms are produced entirely from Born graphs in the
relativistic theory, but integrating out the antinucleons generates a seagull
term in the third-order Lagrangian which has a fixed coefficient\cite{ber92}.
This illustrates a point which we will come back to, namely that one cannot
determine by inspection which graphs in HBCPT are one-particle reducible.
The loop diagrams of Fig.~1 have contributions of order $\omega$ which cancel
and so do not affect the LET, while the $\omega^3$ terms give the
polarisabilities at this order.

\begin{figure}[t]
%\figurebox{20pc}{15pc}{} % to have a box alone
\epsfxsize=27pc % will enlarge or reduce the postscript figures based on the xsize
\epsfbox{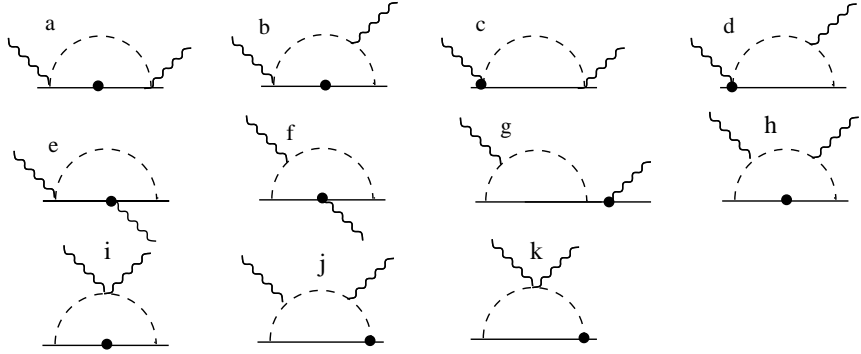} % postscript image file name
\caption{Diagrams which contribute to spin-dependent forward Compton
scattering in the $\epsilon\cdot v=0$ gauge at NLO. The solid dots 
are vertices from ${\cal L}^{(2)}$.}
\end{figure}

At NLO, the diagrams which contribute are given in Fig.~2.  
In the Breit frame, only diagrams 2a-h contribute, and  there can be no
seagulls at this order.  It follows  that there are no undetermined low-energy
constants in the final amplitude.When the  amplitudes are Taylor expanded,
there are  contributions at order $\omega$ and $\omega^3$.  The former do not
violate the LETs, however. The third-order contributions to the LETs actually
involve the bare values of $\kappa$ which enter in the second-order
Lagrangian.  However $\kappa_v$ has a pion loop contribution at the next
order:  $\delta \kappa_v= -g_\A^2 m_\pi M_\N/4 \pi f_\pi^2$.  This then
contributes to the fourth-order Compton scattering amplitude.  Reproducing
these terms is one check on our calculations. The order $\omega^3$ pieces give
the polarisabilities. The requirement $\gamma_5+\gamma_2+2\gamma_4=0$ is
satisfied, which provides another non-trivial check on the results.  

The loop contributions to the polarisabilities to NLO are 
\begin{eqnarray}
\gamma_1&=&{\alpha_{em}g_\A^2 \over 24\pi^2 f_\pi^2 m_\pi^2}\left[1 
-{\pi m_\pi \over 8 M_\N}\bigr(8+5\tau_3\bigr)\right]
\nonumber\\
\gamma_2&=&{\alpha_{em}g_\A^2 \over 48\pi^2 f_\pi^2 m_\pi^2}\left[1 
-{\pi m_\pi \over 4 M_\N}\bigr(8+\kappa_v+3(1+\kappa_s)\tau_3\bigr)\right]
\nonumber\\
\gamma_3&=&{\alpha_{em}g_\A^2 \over 96\pi^2 f_\pi^2 m_\pi^2}\left[1 
-{\pi m_\pi \over 4 M_\N}\bigr(6+\tau_3\bigr)\right]
\nonumber\\
\gamma_4&=&{\alpha_{em}g_\A^2 \over 96\pi^2 f_\pi^2 m_\pi^2}\left[-1 
+{\pi m_\pi \over 4 M_\N}\bigr(15+4\kappa_v+4(1+\kappa_s)\tau_3\bigr)\right]
\nonumber\\
\gamma_0&=&{\alpha_{em}g_\A^2 \over 24\pi^2 f_\pi^2 m_\pi^2}\left[1 
-{\pi m_\pi \over 8 M_\N}\bigr(15+3\kappa_v+(6+\kappa_s)\tau_3\bigr)\right]
\end{eqnarray} 
Although the subleading pieces have a factor of $m_\pi/M_\N$ compared with
the leading piece, the numerical coefficients are often large. The anomalous
magnetic moments are $\kappa_s=-0.12$ and   $\kappa_v=3.71$; with these
values the numerical results for the polarisabilities  to fourth order are
\begin{eqnarray}
\gamma_1&=&[-21.3]  +4.5 -(2.1 +1.3 \,\tau_3)\nonumber\\
\gamma_2&=&         2.3 - (3.1 + 0.7\,\tau_3)\nonumber\\
\gamma_3&=&[10.7]  +1.1 - (0.8 + 0.1\,\tau_3)\nonumber\\
\gamma_4&=&[-10.7]- 1.1  + (3.9 + 0.5\,\tau_3)\nonumber\\
\gamma_0&=&          4.5 -(6.9+1.5\,\tau_3)\nonumber\\
\gamma_\pi&=&[-42.7]+4.5 +(2.7 - 1.1\,\tau_3)
\end{eqnarray}
The term in square brackets, where it exists, is the third-order $t$-channel
pion exchange contribution.  (There is no fourth-order contribution.) 

The NLO contributions are disappointingly large, and call the
convergence of the expansion into question. While the 
fifth-order terms have also been estimated to be large\cite{ber92}, 
this is due to physics beyond $\pi N$ loops, namely the contribution
of the $\Delta$.  Our results show that even in the absence of the $\Delta$,
convergence of HBCPT for the polarisabilities has not yet been reached.

\section{Comments on the definition of polarisabilities}

We now return to the difference between our results and those of the J\"ulich
group, who give expressions for the polarisabilities which are analytically
and numerically different from ours.  The entire difference comes from the 
treatment of diagram 2g, which we include in the polarisabilities and they
omit. The polarisabilities are not in fact usually defined as in Eq.~2, but as
the first term in the expansion of the amplitudes after  subtraction of the
``Born terms".  This removes the LET terms, but, depending on the model
used for the Born graphs, also some $\omega$-dependent terms. 
Gellas \etal argue that the contribution of 2g should also be removed by this 
subtraction. 
 
There are two main objections to this definition.  First, it is not model- and
representation-dependent, as the one-particle reducible part of 2g beyond
the LET piece involves an off-shell ``formfunction" or ``sideways formfactor",
and as stressed by Scherer,\cite{scherer} these cannot be unambiguously defined.
Furthermore the procedeure Gellas \etal have adopted does not respect Lorentz
invariance.  At this order there are terms that vanish in the Breit frame which are 
in fact generated by a lowest-order boost of the third-order (fully irreducible)
loop amplitude.  (In the centre-of-mass frame these show up as pieces with,
apparently, the wrong crossing symmetry: they are even in $\omega$ in amplitudes
$A_3$ to $A_6$, and start at $\omega^4$). However the prescription of Gellas
\etal discards the contribution of 2g to these pieces, violating the boost
invariance of the resulting LO+NLO amplitude. (As Mei\ss ner explained in his
talk, their prescription is to discard the part of 2g which  
has the form $f(\omega)/\omega$, where $f$ is analytic. In fact pieces like this
also arise from other diagrams, notably 2f, while diagrams  2a-e, though
apparently irreducible, contribute LET pieces.  The distinction  between
reducible and irreducible in HBCPT is hidden, as mentioned earlier.)

The other objection to excluding so much from the definition of the
polarisability is that, even if it is done consistently, it does not correspond
to the definition used in the extraction from fixed-$t$ dispersion relations.  There, the
polarisabilities are related to the integral of the imaginary part of the amplitudes
over the cut, where the amplitudes used have effectively been subtracted at the 
point where an intermediate nucleon would be on shell.\cite{babusci}  
This can at most change the spin polarisabilities by something of order
$\alpha_{em}\mn^{-4}$, which is small numerically and is NNLO in HBCPT.

Thus the exclusion of 2g from the ``structure constants" such as polarisabilities
is neither a consistent  definition, nor one that corresponds to dispersion relation 
determinations.

\section*{Acknowledgments}

JMcG and MCB acknowledge the support of the UK EPSRC. VK held a Commonwealth
Fellowship while in Manchester.

\end{document}